# Lévy distribution and long correlation times in supermarket sales

Robert D. Groot

Unilever Research Vlaardingen

PO Box 114, 3130 AC Vlaardingen, The Netherlands

Sales data in a commodity market (supermarket sales to consumers) has been analysed by studying the fluctuation spectrum and noise correlations. Three related products (ketchup, mayonnaise and curry sauce) have been analysed. Most noise in sales is caused by promotions, but here we focus on the fluctuations in baseline sales. These characterise the dynamics of the market. Four hitherto unnoticed effects have been found that are difficult to explain from simple econometric models. These effects are: (1) the noise level in baseline sales is much higher than can be expected for uncorrelated sales events; (2) weekly baseline sales differences are distributed according to a broad non-Gaussian function with fat tails; (3) these fluctuations follow a Lévy distribution of exponent $\alpha$ = 1.4, similar to financial exchange markets and in stock markets; and (4) this noise is correlated over a period of 10 to 11 weeks, or shows an apparent power law spectrum. The similarity to stock markets suggests that models developed to describe these markets may be applied to describe the collective behaviour of consumers.

# 1   Introduction

The search for general trends in markets has a long history. Nevertheless, some 70% of new products put in the market fails. In marketing science the adoption of new technologies is traditionally studied using the Bass model [1], which was first posed in 1969, and which relates the fraction of consumers adopting a new technology per unit of time, to the fraction present at that time. This model ignores the variation among consumers and it ignores their network of social relations. One important recent insight in the description of human economic behaviour is that these networks may be very important for the success or failure of new products. Arthur [2] argued that the amount of information humans typically have deal with is too large to handle, and moreover individuals usually have to base their decisions on partial information. Therefore they turn to rules of thumb, or strategies, to determine their actions. These strategies are often just copied from peers; this is the first reason why social networks are important.

The second reason why networks are important, is that part of the perceived value (or utility) of a product depends on its acceptance in the social network of the consumer. The utility of a product represents more than its direct application; its value in use can be increased because the consumer feels that it attributes to his/her social status [3]. For example, clothing can be valuable to express the membership of a peer-group, or otherwise expensive cars may cue status to the peer-group. This social effect is not limited to products like cars, but may also hold for commodity markets.

Recently Stauffer [4], Weisbuch and Solomon have studied the impact of social networks and percolation phenomena to the adoption of new products. Solomon *et al.* [5] discuss the possibility of observing self-organised criticality. The question addressed is: why does one observe that *some* newly launched products are doomed to failure while *others* become great hits, rather than a featureless distribution of partial success? The general explanation for such a bimodal distribution is discussed in terms of bounded rationality. Decisions are not taken upon complete information, but consumers copy the behaviour of their peers [6]. Thus, the spread of a new technology is often compared to the spread of a virus over the population, and is determined by a percolation transition. Goldenberg *et al.* [7] bring these ideas further and compare them to marketing of consumer goods, namely the introduction of cars and LCD color television sets.

However, the models mentioned are all based on theoretical ideas of how consumers choose, the link to empirical data is weak or missing. Before attempting to describe the dynamics of new market introductions, one should at least be able to describe the dynamics of a stationary market. The first step to develop a realistic model of the collective behaviour of consumers, which takes into account psychological and social factors, should be to collect empirical facts that can be used as a standard to test the models. The development of models for financial and stock markets has benefited greatly from the analysis of price fluctuations [8,9]. Since this was the case for our understanding of stock markets, the obvious first step towards building a model of consumer sales, is to analyse the fluctuations in this market. This should unearth the empirical – stylised – facts of consumer sales. Such an analysis seems to be missing in this field. Once the stylised facts are established we can analyse their causes, and finally build models based on these insights. As a first attempt in this direction the fluctuations in actual sales data of a number of consumer goods are analysed. Here we restrict to presenting the empirical data and its implications for modelling.

## 2  The consumer goods market

The available observations are actual sales data of products in supermarkets. Time series of sales can be obtained from ACNielsen [10]. The longest series available is over a period of 120 weeks. As an example of sales in a commodity market, the sales data of ketchup in the Netherlands will be discussed first. In the Netherlands there are about $16 \times 10^6$ consumers, divided over some $7 \times 10^6$ households. These households shop at some 5200 supermarkets. The raw sales data for Calvé, Gouda's Glorie, Heinz and Remia are shown in Figure 1a. Most of the noise in the sales data shown in Figure 1a can be attributed to promotional actions. ACNielsen also provides a baseline for the sales of each product, which is the total sales not conducted under promotion. For the shops where a promotion is on, baseline sales from the previous eight weeks are used to extrapolate to the value one would have had if there would not have been a promotion in that week. Basically it comes down to subtracting the excess sales that is attributed to promotion. The output of this analysis is known as the baseline sales. The noise in the baseline sales is shown in Figure 1b, this is the underlying fluctuation of the market.

The principal parameter to characterise the sales data is the market share of the various brands. In this case these are 52.7%, 19.6%, 14.8% and 12.9% respectively in order of size. As a first method of analysis the baseline sales data in Figure 1b were fitted to parabolas, and the variations from the trends were studied. The root mean square fluctuations thus found vary from 7.6% of sales for the product of highest market share to 13.2% of sales for the product of lowest market share.

The relative noise is an indication of the number of uncorrelated sales events. If we have N independent events the *relative* noise is given by $1/\sqrt{N}$. Hence, from the relative noise in sales we find in turn the number of independent sales events. The remarkable result is that out of 37000 bottles of Heinz ketchup sold per week, apparently only 173 events are independent, and out of 9000 bottles of Remia sold per week only 57 events are uncorrelated. This difference cannot be explained on the basis of the size of the sample taken by ACNielsen. The sample is allegedly based on some 1100 shops (out of 5161), which represents 40% of all sales in the Netherlands; the statistical variation is estimated as 2% [11]. Nor can it be explained by assuming that the data reflects sales to the retailer in larger quantities. The data is actual point of sales data to the consumer [11]. Even if we assume that the number of 1100 shops is overstated (some say that the size of the sample is only 350 shops) then still the noise level does not fit the number of independent buying events.

Hence, merely by studying the variance in baseline sales fluctuations we arrive at the conclusion that the buying behaviour of thousands of consumers is correlated. Whereas this may be an obvious social effect for fashionable products like cloths, it is less obvious why this would be the case for a commodity like ketchup. Therefore we will study the data in more detail in the next sections to obtain the noise spectrum and correlation functions.

## 3   The Noise Distribution

To analyse the data set further, the noise distribution of sales is studied. Mantegna and Stanley introduced a method to analyse the fluctuations in the S&P 500 index of stock markets, by studying the price difference of a stock at two successive points in time (the price return) [8]. In stock markets, this time difference can be as small as one minute, but it can also be days. They found that the distribution of price differences follows a characteristic shape, the Lévy distribution. The width of the distribution obviously depends on the time interval studied, but the shape of the distribution remains the same. For stock markets it is given by

$$P(x) = \frac{1}{\pi} \int_0^\infty \exp(-\gamma q^\alpha) \cos(qx)\, dq \qquad 1$$

where $\alpha$ takes on the value $\alpha = 1.4$ in real markets. For $\alpha = 2$ this equation generates a Gaussian distribution, but for the lower value $\alpha = 1.4$ pertinent to the market, the distribution has fat tails. For large values of the price return the distribution crosses over to a characteristic power law with $\alpha = 3$, thus following a truncated Lévy distribution [12,13].

For ketchup sales data we can use the same method of analysis; the distribution of the change of sales from one week to the next is studied, i.e. the distribution of $\Delta s(t) = s(t+1) - s(t)$. Because we have only limited data, the mean value of $\Delta s$ was first determined for each series, which was subsequently subtracted from the data. Next the spread in $\Delta s$ was determined, and $\Delta s$ was normalised to its spread for each series. This gives four series of zero mean and spread one, with a total of 468 observations (some data points corresponding to Xmas sales were removed). For this data set a distribution has been determined, which is shown in Figure 2.

This figure shows that the distribution of sales fluctuations is *not* a Gaussian distribution. In fact the Lévy distribution taken at the power $\alpha = 1.4$ seen in stock markets [8] makes a very good fit to the ketchup data, whereas the Gaussian distribution of the same width makes a poor fit. The remarkable take out is that for the ketchup market, daily supermarket sales corrected for promotional actions, follows roughly the pattern of price return fluctuations of stock markets. This *suggests* a similar underlying cause. Naively one would expect a Gaussian distribution, as the sales are the result of many independent buying events. When a distribution is not Gaussian, this implies that there are strong correlations in the signal. In other words, the buying events are not independent.

One may wonder if the effect shown here is an artefact, caused by the procedure followed by ACNielsen to obtain the baseline. This is not the case. If we simply subtract all sales in shops where promotional actions are on and then study the fluctuations, we find the same distribution. Further, one may wonder if the broad noise spectrum may be a remnant of promotional sales. Perhaps a small residual error in subtracting the promotional sales has led to the Lévy noise. If this would be the case the effect should disappear when we take out all high frequency modes, since the promotions are spiky one-week events. To this end the data was Fourier transformed, and all modes of frequency $\omega > 1$ were removed from the sales data (i.e. all frequencies with a period smaller than 6.3 weeks are taken away). The back-transform for each brand is now a smooth curve running through the baseline sales. When a distribution of sales differences from one week to the next is calculated from this smooth curve and scaled relative to the noise amplitude, we find again a Lévy distribution (see Figure 3), and within the statistical accuracy this is indistinguishable from the previous distribution. This shows that the broad distribution and its fat tails are *not* caused by high frequency noise, since everything with a period less than 6 weeks has been taken away. This implies that the shape of this distribution is *not* caused by promotions, *not* caused by weather variations and *not* caused by isolated sale bursts at bank holidays.

Finally, if this effect is no coincidence, one should be able to reproduce it for other markets. Products deemed to be similar to ketchup are mayonnaise and curry sauce, in that they are all sauces, and not bought very frequently. We studied the baseline sales of Calvé, Gouda's Glorie and Remia mayonnaise, and of Calvé, Gouda's Glorie, Hela, Heinz, Remia and AO curry sauce over the same time interval as the ketchup data. Baseline sales in these markets indeed look as noisy as that of ketchup.

For the mayonnaise and curry baseline sales data, again the weekly sales differences were determined. For each brand and product, the sales differences were then normalised to the mean spread over the 118 week period (again Xmas 2000 sales was excluded from the statistics). When the mayonnaise and curry sauce markets are studied separately the same distribution is obtained as for ketchup. All three markets are therefore statistically the same as far as the noise distribution is concerned. Because each data series is normalised to unit variance, all ketchup, mayonnaise and curry data can be collected into one distribution. This distribution matches the Lévy distribution close to exactly, see Figure 4. Again, the Lévy exponent 1.4 – copied from financial markets – fits the data.

## 4 Correlation functions

One may argue that the shape of the distribution and the high noise level could be caused by a cross correlation effect between promotions of one brand and the sales of another. The idea behind this is that baseline sales data of a brand has been corrected for promotions of that same brand, but not for promotions of other brands. If brand 1 has a promotion leading to sales increase of brand 1, then brands 2, 3 and 4 might suffer as a result. Thus, one might expect a negative cross correlation between the sales of one brand and the baseline sales of another. If this is true, the noise level observed in the previous section may not be caused by the dynamics of the market, but could still be a remnant of promotion effects. To check this hypothesis the auto-correlation and cross-correlation functions were studied.

If the weekly sales of brand $i$ is given by $s_i(t)$, the sales correlation is given by

$$C_{ij}(t) = \frac{1}{\sqrt{\overline{s_i}\,\overline{s_j}}\,(T-t)} \int_0^{T-t} (s_i(t') - \overline{s_i})(s_j(t'+t) - \overline{s_j})\, dt' \qquad 2$$

where $^{—}$ stands for a time average. Since the variance should be proportional to the number of sales events, $C_{ij}(t)$ should be independent of the size of the brand. This is checked for all sauces (ketchup, mayonnaise and curry cause) and indeed found to be the case. We first concentrate on the baseline sales auto-correlation, and then study cross correlations. The value of the auto-correlation at $t = 0$ is the number of sales events that are correlated. To see this, let mean sales be given by $<s> = Ns_0$ where N is the number of independent events, and $s_0$ the average amount sold in such an event. The variance in the mean is $\sigma^2 = Ns_0^2$, hence $C(0) = \sigma^2/<s> = s_0$.

If the broad noise distribution is a remnant of promotions, one would expect a sharp peak at $t = 0$ and no signal after the first week, since promotions only run for one week. To calculate the sales correlation functions, the data can be detrended or not. Detrending means that the mean trend over the 118-week period is subtracted from the sales curves, such that the mean slope will vanish [14]. This obviously takes away fluctuations on the time scale of two years or more, but it would correct for a continually rising or declining market. For ketchup, detrending or not detrending has little influence on the result, for curry sauce and mayonnaise the differences are larger. Taking the mean and half difference of the detrended and non-detrended correlation functions as our estimate and its error bar, the results for all three markets are within the uncertainty of the others. Thus, within a first approximation the three markets show the same sales auto-correlation function, which is shown in Figure 5.

The small dots in Figure 5 are brand averages for the individual categories; the big squares with error bars represent the average and spread over the three markets. Two things should be noted. Firstly, in all markets the point at $t = 0$ lies above the extrapolated value determined from the other points. The actual values at $t = 0$ are $C_{ii}(0) = 206$, 206 and 340 for ketchup, curry sauce and mayonnaise, whereas the extrapolated values respectively are given by 118±7, 122±5 and 234±8. For the average over all products we find an actual value at $t = 0$ of $C(0) = 251$, and an extrapolated value of 158±4. The error bars indicated are the uncertainty in the extrapolation, obtained by fitting an exponential through the correlation function. This does not include the systematic error that shows up in the difference between detrending the data or not. This error is similar in size to the error bars between the markets, which are shown in Figure 5.

Indeed, the auto-correlation function in $t = 0$ is higher than the value extrapolated from the correlation at $t > 0$. This excess can be attributed to be a remnant of the promotions, but this represents only 37% of the variance. Other effects like weather variations or holidays may have also an influence on this term. Most of the variance is related to fluctuations of the baseline, and is thus caused by intrinsic market moves. As can be seen from Figure 5, these fluctuations have a very long correlation time as compared to the typical 4 minutes seen in stock markets [12]. The mean correlation function averaged over all brands and markets for $t > 0$ is given by

$$C_{ii}(t) \approx (158 \pm 4)\exp(-\frac{t}{10.2 \pm 0.4}) \qquad 3$$

where $t$ is in weeks. The slow correlation time points at an internal dynamical process, this could be customer loyalty and repeat buys, but certainly cannot be due to promotions.

We now discuss the cross correlations, and focus upon the cross-correlations between ketchup brands. When the top brand has a large peak due to a promotional action one might expect the other brands to have a (small) sales decrease. This, however, is not seen. Whether we correlate the full sales with baseline sales, or take the correlation between baseline sales does not matter significantly for the result. The cross correlation function between top brand and the others vanishes within the noise, although close to $t = 0$ a small *positive* correlation might be present, see Figure 6. This observation again confirms that because the cross correlation is zero, the noise in the baseline cannot be a remnant of the promotional noise. Figure 6 also shows the mean cross correlation between the three lower ketchup brands. Instead of having a negative correlation at $t = 0$, it has an unexpected *positive* correlation. The amplitude at $t = 0$ is $C_{ij}(0) = 69\pm3$, and the decay time is $\tau$ = 10.9$\pm$0.8 weeks. Within the uncertainty, the cross-correlation time between the lower brands is thus equal to the auto-correlation time. Hence the whole market is governed by a process with a single time scale of about 10 - 11 weeks.

To gain further insight into the nature of the processes that determine the baseline sales dynamics, the Fourier transform of the baseline sales data, $s(\omega) = 2/T \int_0^T s(t) \, exp(i\omega t) \, dt$ is studied. If indeed there are two processes, one of which is much faster than the other, this should show up as a signal at high frequency. Thus we could separate the high frequency noise from the noise spectrum of the slow process. The baseline sales correlation is readily obtained as

$$C(\omega) = \frac{1}{N}\sum_{i=1}^{N} s_i(\omega)s_i(-\omega)/<s_i> \qquad 4$$

$C(\omega)$ is actually the Fourier transform of the function $C_{ii}(t)$ defined above, and is an average over all brands. There is a subtle difference though. By taking the Fourier transform the convolution is calculated, implicitly assuming that the function s is periodic in T, whereas in Eq 2 the time average is restricted to the interval where t+t' < T, and no periodicity is assumed. The error bars are the estimated error in the mean, based on the spread in the data for each point. The autocorrelation functions of the baseline sales have been determined for 11 brands and products. Two curry brands had long running new product introductions, which were counted as promotions by ACNielsen. This seriously affects the baseline for these brands and therefore these were left out of the statistics. If we average over all 11 products and assume that there is a unique mean value at each frequency, we obtain the function shown in Figure 7. The points are averaged over two to four successive frequencies to reduce noise. The logarithm of the correlation function is averaged over all brands. The full curve shown in Figure 7 is a fit to a single Maxwell mode and a constant noise term, $G(\omega) = a/(1+(\omega\tau)^2)+c$, the dashed curve is a fit to a power law plus a constant, $G(\omega) = 1/(\omega\tau)^\alpha+c$. Within the uncertainty of the data the two fit functions are equally valid. Hence, over about a decade in frequency the correlation function scales with an apparent power law $G(\omega) \propto \omega^{-\alpha}$ with power $\alpha = 1.5\pm0.1$.

## 5  Discussion

The analysis of baseline sales of three different products (ketchup, mayonnaise and curry) has led to four new observations, namely:

1. the sales noise level is much higher than can be expected for uncorrelated buying events
2. weekly sales differences show a broad non-Gaussian distribution with fat tails
3. this sales distribution is similar to the price return distribution of stock markets
4. the baseline auto-correlation shows a long correlation time or an apparent power law behaviour at low frequency.

A trivial explanation for the first observation might be that the sample upon which the sales data is based is actually smaller than claimed. However, this explanation can be ruled out by the data. The correlation function averaged over all products extrapolates to zero at a mean value of about 160. This means that the intrinsic variance in the baseline sales is 160 times larger than actual sales. If this is to be explained on the basis is the sample size, the data must be based on only 32 shops. This is not realistic. Moreover, this trivial explanation would imply that the noise level is based on fewer sales events that are all independent. Such a model of independent consumers, going through independent buying decisions, would directly lead to Gaussian fluctuations. This again is falsified by the data: the fluctuation spectrum resembles that of a (truncated) Lévy flight. This points at strong correlations between the behaviour of consumers. With other words, the notion of independent consumers who take independent decisions is false.

RD Groot, Lévy distribution and long correlation times in supermarket salesThe explanation that the noise is a remnant of promotions is also ruled out by the data. Indeed the extrapolated value of the auto-correlation function is lower than the actual value at t = 0. The difference can be attributed to short time effects like promotions, weather etc, but this can only explain some 37% of the variance. The major part of the fluctuations is related to an intrinsic process on a time scale of some 10 weeks.

Lévy distributions have been observed recently in commodity markets [15], but unlike the earlier result by Mandelbrot for the cotton market [16], these studies showed higher exponents of distributions that are not Lévy stable. The reason for this is that the market crosses over from Lévy stable behaviour with $\alpha \approx 1.4$, to power law behaviour with $\alpha \approx 3$. The crossover point is located near an absolute return of roughly 3 standard deviations [9,12,15]. Mandelbrot did not have enough data to see this crossover. The present analysis of sales to consumers also lacks the amount of data to see this limiting law. However, the result in Figure 4 is consistent with a truncated Lévy distribution that crosses over by 3 standard deviations. It should be mentioned here that weekly sales returns of soups have also been analysed, and added to the distribution. For this combined data set we do start to see a crossover. However, soup consumption is very season dependent. Hence, to analyse this data the yearly variation had to be subtracted by detrending against periodic functions. This also led to spurious oscillations in the auto-correlation function. For this reason we cannot say whether this crossover in the combined data set is fortuitous or not, hence the data was left out of Figure 4.

Unlike for stock market price returns, the auto-correlation function of baseline sales fluctuations shows a long time correlation. Whether this correlation is an exponential decay or a power law cannot be determined with complete certainty. The analysis in real time (Figure 5 and Figure 6) suggests an exponential, but the size of the error bars should be taken seriously. This is a systematic error that shows the difference between detrending and not detrending the data. This strong influence implies that there are modes on the time scale of the full 120-week period. This result is qualitatively in line with the recent analysis by Sornette *et al* of endogenous versus exogenous shocks in complex networks, who find power law decay in peaks of Amazon book sales [17,18].

There are basically two different explanations for the occurrence of fat tails in markets. The first is that the system is in a self-organised critical state [19], and that "avalanches" occur regularly. The second explanation is that actions are correlated because agents are linked in a social network. The first explanation is closely related to the theory of avalanche dynamics. In this theory 1/f-noise is related to anomalous random walks of the underlying process [20].

A class of deterministic models introduced by Olami, Feder and Christensen exhibits 1/f-noise with an exponent that depends on the degree of conservation [21]. In a similar model of artificial societies a cellular automaton evolves towards a critical state, with avalanches obeying finite size scaling [22]. This principle of self-organized criticality has been applied to financial market modelling by Ponzi and Aizawa [23]. Their model describes an evolving network of traders, subject to Darwinian evolution. It produces a Lévy distribution of the correct exponent for the price returns, and it reproduces clustered volatility.

The simplest market model exhibiting self-organised criticality is the Minority Game. This has been applied to market simulation with different species of traders with finite capital [24]. Wealth is accumulated in the course of the simulation, and the richer agents have more influence on the market. As in the standard Minority Game, a phase transition is found, depending on the ratio of the information complexity and the number of agents. The emergent picture is that when agents are few, the market is rich of profitable opportunities. These attract others to the market. As the number of agents increases, the opportunities are eliminated and the market is driven towards its critical point. This suggests why real markets operate close to the critical point where profitable trade opportunities are barely detectable. Challet *et al* [24] state that the process by which the market self-organises is likely to be of evolutionary nature, and hence takes place on longer time-scales.

There is an obvious competition amongst the *producers*, which may give rise to an evolutionary effect (including "avalanches"), but whether or not this can explain the fast sales variations on a weekly basis is questionable. Evolution is related to product improvements, and their fitness to the market. Indeed the market is very dynamic, and new products are introduced regularly, but the typical time scale involved is months rather than weeks. One may speculate that this corresponds to the 10-11 week time scale seen in the sales auto-correlation function. On the other hand, in the examples studied here only three to six leading brands are competing, which is a small number for such complex behaviour, even if each brand actually holds six or more different items in their product portfolio.

The second explanation for stylised fact in markets is that the behaviour of agents is correlated because they are linked in a social network. If they tend to follow the decisions of their peers, sales becomes correlated which gives rise to larger sales fluctuations. A simulation model that incorporates this effect must contain a large number of consumer groups, that have mutual interactions and that drive the market system to a critical point. In marketing language, these groups may correspond to the particular segmentation of the market. The critical point should then correspond to a consumer segment distribution that is given by Zipf's law [13]. The archetypal model to describe such correlated behaviour due to social networks is the Cont-Bouchaud percolation model [25]. This model is based on a random communication structure between agents, which gives rise to power-laws in price fluctuations and herding behaviour. The model assumes that agents form coalitions, where all agents in one coalition take the same (random) decision to buy, sell, or do nothing. The coalitions are formed through a stochastic process, where each agent is linked to any other one with a given probability. At a critical probability, the system shows a percolation transition, where a very broad distribution of cluster sizes appears. Since the cluster size distribution has an algebraic decay, so does the distribution of price variations because the price variation follows the excess demand. Hence, the correct stylised facts are found when the correct cluster size distribution is put in by hand [13,26], or when the random clusters are formed by simulating the Ising model at its critical point [27]. Lux and Marchesi describe a multi-agent model of financial markets which supports the idea that scaling arises from mutual interactions of participants. Their model contains a news arrival processes that lacks power-law scaling and time dependence in volatility, yet it generates such behaviour as a result of interactions between agents [28]. Finally, social networks are key in the analysis of Sornette *et al* of endogenous and exogenous shocks in book sales [17,18].

In reality both the competitive evolution of brand fitness *and* the network structure of the consumers may be pertinent. However, there are two arguments that point to the latter effect as being dominant. Firstly, the avalanches seen in evolution theory describe the correlated *extinction* of many species. Translated to the market this would mean many brands going bust at the same time, rather than many consumers buying the same product at the same time. This is not in line with observations. Secondly, 58% of the long time variance in ketchup sales is cross-correlated between brands, and the time scales for the auto-correlation and cross-correlation functions are the same. This points at a cause for the fluctuations that is common to all brands, rather than to the brand management strategy and evolution of brands. The question thus arising is how these clusters of correlated consumer behaviour are formed. Why do all consumers simultaneously decide to buy mayonnaise? It is speculated that advertisements play a crucial role in forming such correlation, and in generating the non-Gaussian distribution of sales returns, as this may create a broad distribution of consumers who have and who have not been influenced.

## Acknowledgements

The author thanks Dr PAD Musters for providing data and for critical discussions.

## Figure captions

**Figure 1** a, Raw sales data for four brands of ketchup in the Netherlands; b, ketchup baseline sales.

**Figure 2** Distribution of fluctuations in ketchup baseline sales, averaged over four brands. The full curve is a fit to the Lévy distribution, the dashed curve is a Gaussian distribution of the same width.

**Figure 3** Weekly sales difference distribution for raw baseline sale (circles and dashed curve) and Fourier filtered time series (black dots and full curve).

**Figure 4** Distribution of weekly sales differences averaged over 13 products and brands. The full curves are fits to a Lévy distribution of exponent $\alpha$ = 1.4.

**Figure 5** Mean auto-correlation of baseline sales of three markets: ketchup (dots), curry sauce (circles) and mayonnaise (triangles). The squares with error bars represent the average and spread over all three markets. The straight line is an exponential with decay time 10 weeks.

**Figure 6** Cross-correlation between ketchup market leader sales and sales of other brands (lower curve), and between the smaller brands mutually. No correlation is apparent between top brand and lower brands, but the lower brands are mutually correlated.

**Figure 7** Auto-correlation function, averaged over all brands and products. Closed dots: non-detrended data, open circles: detrended data. The full curve is a fit to a single Maxwell mode plus a constant, the dashed curve is a power law plus constant.

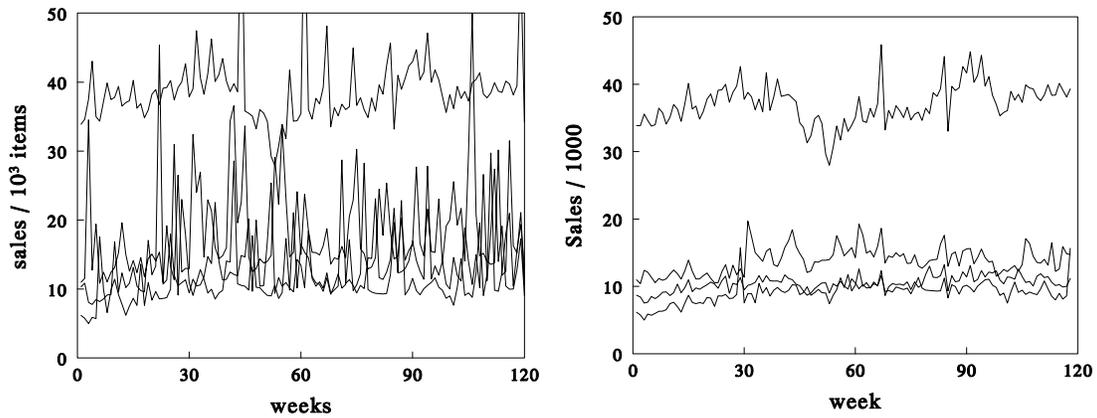

**Figure 1** a, Raw sales data for four brands of ketchup in the Netherlands; b, ketchup baseline sales.

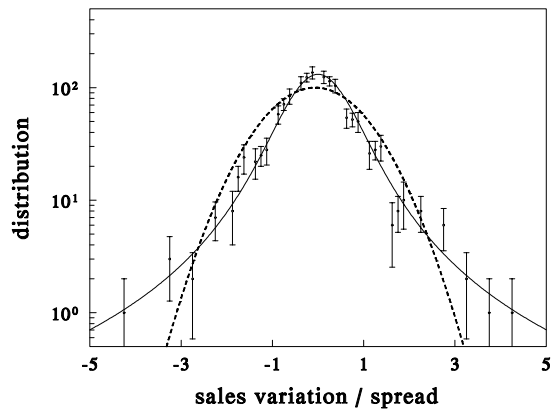

**Figure 2** Distribution of fluctuations in ketchup baseline sales, averaged over four brands. The full curve is a fit to the Lévy distribution, the dashed curve is a Gaussian distribution of the same width.

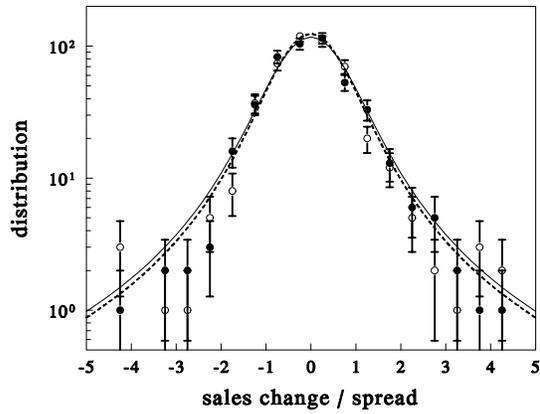

**Figure 3** Weekly sales difference distribution for raw baseline sale (circles and dashed curve) and Fourier filtered time series (black dots and full curve).

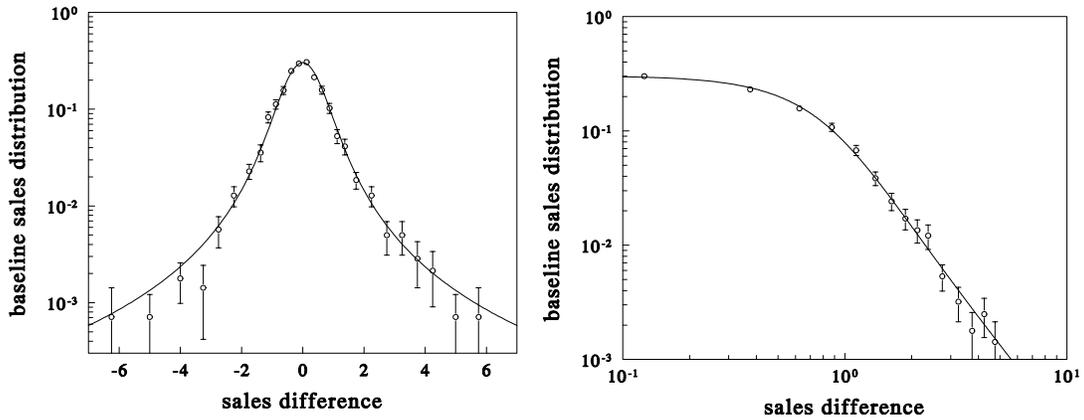

**Figure 4** Distribution of weekly sales differences averaged over 13 products and brands. The full curves are fits to a Lévy distribution of exponent $\alpha = 1.4$.

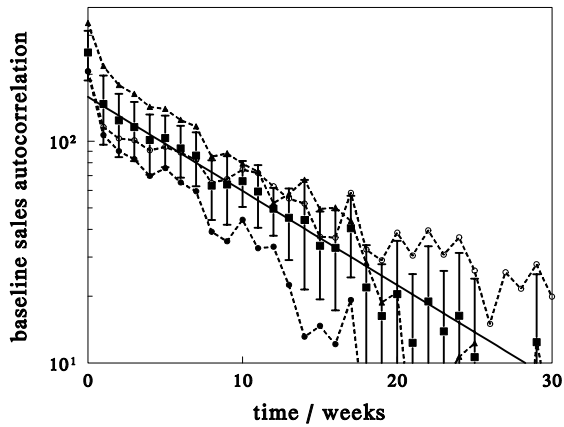

**Figure 5** Mean auto-correlation of baseline sales of three markets: ketchup (dots), curry sauce (circles) and mayonnaise (triangles). The squares with error bars represent the average and spread over all three markets. The straight line is an exponential with decay time 10 weeks.

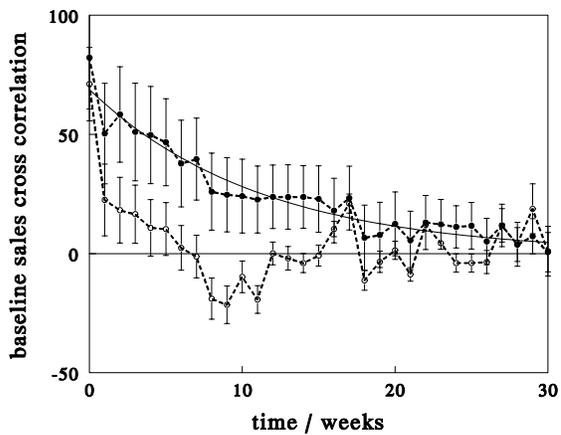

**Figure 6** Cross-correlation between ketchup market leader sales and sales of other brands (lower curve), and between the smaller brands mutually. No correlation is apparent between top brand and lower brands, but the lower brands are mutually correlated.

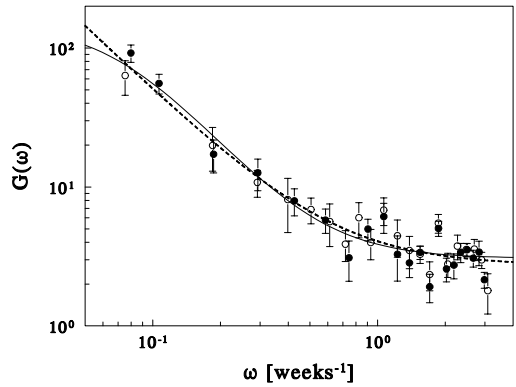

**Figure 7** Auto-correlation function, averaged over all brands and products. Closed dots: non-detrended data, open circles: detrended data. The full curve is a fit to a single Maxwell mode plus a constant, the dashed curve is a power law plus constant.